\documentclass[11pt]{article}
\usepackage{amsmath}

\usepackage{subfig}
\usepackage{color}
\usepackage{amssymb}
\usepackage{multirow}
\usepackage{natbib}
\usepackage{graphicx}
\usepackage[dvips]{epsfig}
\usepackage{lingmacros}
\newcommand{\be}{\begin{eqnarray}}
\newcommand{\ee}{\end{eqnarray}}
\newcommand{\ba}{\begin{eqnarray*}}
\newcommand{\ea}{\end{eqnarray*}}
\newtheorem{theorem0}{Theorem}
\newtheorem{lemma0}{Lemma}
\newtheorem{remark0}{Remark}
\newtheorem{fact0}{Fact}
\newtheorem{example0}{Example}
\newtheorem{corollary0}{Corollary}
\newtheorem{proposition0}{Proposition}
\newtheorem{conjecture0}{Conjecture}

\def\boldfacefake #1{%
    \hbox{%
        \mathsurround=0pt
        \hbox to 0.4pt{$#1$\hss}%
        \hbox to 0.4pt{$#1$\hss}%
        \hbox {$#1$}%
    }%
}
\newcommand{\nmathbf}{\bf}

\def\bfI{\mbox{$\nmathbf I$}}

\def\bfV{\mbox{$\nmathbf V$}}

\def\bfx{\mbox{$\nmathbf x$}}

\newcommand{\go}{\rightarrow}

\newcommand{\expect}{\mbox{\rm I\kern-.20em E}}
\newcommand{\reals}{\mbox{\rm I\kern-.20em R}}
\newcommand{\sreals}{\mbox{\small \rm I\kern-.20em R}}

\newcommand{\cov}{\mbox{cov}}
\newcommand{\var}{\mbox{var}}

\newcommand{\qed}{\nobreak \ifvmode \relax \else
      \ifdim\lastskip<1.5em \hskip-\lastskip
      \hskip1.5em plus0em minus0.5em \fi \nobreak
      \vrule height0.75em width0.5em depth0.25em\fi}

\begin{document}
\title{\bf A new three-parameter lifetime distribution and associated inference}
\author{Min Wang\thanks{Corresponding author. Email: minwang@mtu.edu}\\
 {\small
  Department of Mathematical Sciences, Michigan Technological University, Houghton, MI 49931, USA}
   }
\date{}        
\maketitle

\begin{abstract}

In this paper, a new three-parameter lifetime distribution is introduced and many of its standard properties are discussed. These include shape of the probability density function, hazard rate function and its shape, quantile function, limiting distributions of order statistics, and the moments. The unknown parameters are estimated by the maximum likelihood estimation procedure. We develop an EM algorithm to find the maximum likelihood estimates of the parameters, because they are not available in closed form. The Fisher information matrix is also obtained and it can be used for constructing the asymptotic confidence intervals. Finally, a real-data application is given to demonstrate the performance of the new distribution. 

\textbf{Keywords}: Compounding; Lindley distribution; geometric distribution; maximum likelihood estimation; EM algorithm; lifetime distribution

\textbf{2000 MSC}: 62F10, 60E05, 62P99
\end{abstract}

\section{Introduction} \label{section:01}

Suppose that a company has $N$ systems functioning independently and producing a certain product at a given time, where $N$ is a random variable, which is often determined by economy, customers demand, etc. The reason for considering $N$ as a random variable comes from a practical viewpoint in which failure (of a device for example) often occurs due to the present of an unknown number of initial defects in the system. In this paper, we focus on the case in which $N$ is taken to be a geometric random variable with the probability mass function given by
\begin{equation*}
P(N = n) = (1-p)p^{n-1},
\end{equation*}
for $0 < p < 1$ and $n = 1, 2, \cdots$. Note that $N$ can also be taken to follow other discrete distributions, such as binomial, Poisson, etc, whereas they need to be truncated 0 because one must have $N \geq 1$. \textit{Another rationale by taking $N$ to be a geometric random variable is that the ``optimum'' number can be interpreted as ``number to event'', matching up with the definition of a geometric random variable}, as commented by \cite{Nada:Canc:Orte:2013}. Other motivations can also be found in \cite{Nada:Canc:Orte:2013}. In fact, the geometric distribution has been widely used for the number of ``systems'' in the literature; see, for example, \cite{Conti:Greg:2000}, \cite{Fric:Gast:Moha:2012}, to name just a few. In addition, the geometric distribution has been adopted to obtain some new class of distributions for modeling lifetime data. Among others, we refer the interested readers to \cite{Adam:Louk:1998} for the exponential geometric (EG) distribution, \cite{Reza:Nada:Tahg:2011} for the exponentiated exponential geometric (EEG) distribution, \cite{Nada:Canc:Orte:2013} for the geometric exponential Poisson (GEP) distribution, and references cited therein.

On the other hand, we assume that each of $N$ systems is made of $\alpha$ parallel components, and therefore, the system will completely shutdown if all of the components fail. Meanwhile, we also assume that the failure times of the components for the $i$th system, denoted by $Z_{i1}, \cdots, Z_{i\alpha}$, are independent and identically distributed (iid) with the cumulative distribution function (cdf) $G(z)$ and the probability density function (pdf) $g(z)$. For simplicity of notation, let $Y_i$ stand for the failure time of the $i$th system and $X$ denote the time to failure of the fist out of the $N$ functioning systems, that is, $X = \min (Y_1, \cdots, Y_N)$. Then the conditional cdf of $X$ given $N$ is given by
\begin{align*}
G(x \mid N) &= 1 - P(X > x \mid N)= 1 -P^N(Y_1 > x)\\[+3pt]
&= 1 -\bigl[1 - P(Y_1 < x)\bigr]^N = 1 -\bigl[1 - P^\alpha(Z_{11} < x)\bigr]^N\\[+3pt]
&= 1 -\bigl[1 - G(x)^\alpha\bigr]^N,
\end{align*}
and the unconditional cdf of $X$ can thus be written as
\begin{align} \label{cdf:01} \nonumber
F(x) &= \sum_{n=1}^\infty  G(x \mid N)P(N=n)\\[+3pt] \nonumber
     &= \sum_{n=1}^\infty\Bigl\{ 1 - \bigl[1 - G(x)^\alpha \bigr]^n\Bigr\}P(N=n)\\[+3pt] \nonumber
     &= 1 - (1-p) \bigl(1 - G(x)^\alpha\bigr)\sum_{n=1}^\infty \bigl[p(1-G(x)^\alpha)\bigr]^n\\[+3pt]
     &= \frac{G(x)^\alpha}{1 - p + p\cdot G(x)^\alpha}.
\end{align}

The new class of distribution (\ref{cdf:01}) contains several lifetime distributions as special cases. As an illustration, if the failure times of the components for the $i$th system are iid exponential random variables with scale parameter $\beta$, that is, $G(z) = 1 - e^{-\beta z}$, then we obtain the EEG distribution due to \cite{Reza:Nada:Tahg:2011}. Its cdf is given by
\be
F(x) = \frac{\bigl(1 - e^{-\beta x}\bigr)^\alpha}{1 - p + p\bigl(1 - e^{-\beta x}\bigr)^\alpha}.
\ee
Note that in reliability engineering and lifetime analysis, we often assume that the failure times of the components within each system follow the exponential lifetimes; see, for example, \cite{Adam:Louk:1998}, \cite{Reza:Nada:Tahg:2011}, among others. This assumption may seem unreasonable because for the exponential distribution, the hazard rate is a constant, whereas many real-life systems do not have constant hazard rates, and the components of a system are often more rigid than the system itself, such as bones in a human body, balls of a steel pipe, etc. Accordingly, it becomes reasonable to consider the components of a system to follow a distribution with a non-constant hazard function that has flexible hazard function shapes.

In this paper, we introduce a new three-parameter lifetime distribution by compounding the Lindley and geometric distributions based on the new class of distribution (\ref{cdf:01}). The Lindley distribution was firstly proposed by \cite{Lind:1958} in the context of Bayesian statistics, as a counterexample of fiducial statistics. It has not been very well explored in the literature partly due to the popularity of the exponential distribution in statistics, especially in reliability theory. Nonetheless, it has recently received considerable attention as an appropriate model to analyze lifetime data especially in applications modeling stress-strength reliability; see, for example, \cite{Zake:Dola:2009}, \cite{Mazu:Achc:2011}, \cite{Gupt:Sing:2012}. Recently, \cite{Ghit:Atie:Nada:2008} argue that the Lindley distribution could be a better lifetime model than the exponential distribution through a numerical example. In addition, they show that the hazard function of the Lindley distribution does not exhibit a constant hazard rate, indicating the flexibility of the Lindley distribution over the exponential distribution. These observations motivate us to study the structure properties of the distribution (\ref{cdf:01}) when the failure times of the units for the $i$th system are iid Lindley random variables with parameter $\theta$, that is
\be \label{lind:cdf}
G(z) = 1 - \frac{\theta +1 + \theta z}{\theta +1} e^{-\theta z}, \quad z >0,
\ee
where the parameter $\theta >0$. The corresponding cdf of the new distribution is defined by
\be \label{new:cdf}
F(x) = \frac{\Bigl(1 - \frac{\theta +1 + \theta x}{\theta +1} e^{-\theta x} \Bigr)^\alpha}{1 - p + p\Bigl(1 - \frac{\theta +1 + \theta x}{\theta +1} e^{-\theta x} \Bigr)^\alpha}, \quad x>0,
\ee
where the parameters $\alpha >0$, $\theta >0$, and $0 < p <1$. We refer to the distribution given by (\ref{new:cdf}) as the \textit{exponentiated Lindley geometric} (ELG) distribution. The main reasons of introducing the ELG distribution can be summarized as follows. (i) The ELG distribution contains several lifetime distributions as special cases, such as the Lindley-geometric (LG) distribution due to \cite{Zake:Mahm:2012} for $\alpha=1$. (ii) It is shown in Section \ref{sub:mom} that the ELG distribution can be viewed as a mixture of exponentiated Lindley distributions introduced by \cite{Nada:Bako:2011}. (iii) The ELG distribution is a flexible model which can be widely used for modeling lifetime data. (iv) The ELG distribution exhibits monotone as well as non-monotone hazard rates but does not exhibit a constant hazard rate, which makes the ELG distribution to be superior to other lifetime distributions, which exhibit only monotonically increasing/decreasing, or constant hazard rates. (v) The ELG distribution outperforms several of the well-known lifetime distributions with respect to a real-data example.

The remainder of the paper is organized as follows. In Section \ref{section:02}, we investigate various properties of the new distribution, including shape of the pdf, hazard rate function and its shape, quantile function, limiting distributions of order statistics, and the $n$th moments. Estimation using the maximum likelihood procedure is discussed in Section \ref{section:03}, and an EM algorithm is proposed to find the maximum likelihood estimates because they cannot be obtained in closed form. In Section \ref{section:04}, a real-data application is given to illustrate the superior performance of the ELG distribution over several well-known lifetime distributions. Some concluding remarks are given in Section \ref{section:05}.

\section{Properties of the distribution} \label{section:02}

In this section, we provide various mathematical properties of the ELG distribution. These include the pdf and its shape (Section \ref{sec:01}), hazard rate function and its shape (Section \ref{sec:02}), quantile function (Section \ref{sec:03}), limiting distributions of order statistics (Section \ref{sec:04}), and expressions for the $n$th moments (Section \ref{sub:mom}).

\subsection{Probability density function} \label{sec:01}

The corresponding pdf of the ELG distribution corresponding to (\ref{new:cdf}) is given by
\be \label{pdf:01}
f(x) = \frac{\alpha\theta^2(1-p)(1+x)e^{-\theta x}\Bigl(1 - \frac{\theta +1 + \theta x}{\theta +1} e^{-\theta x} \Bigr)^{\alpha-1}}{ (\theta+1)\biggl[1 - p + p\Big(1 - \frac{\theta +1 + \theta x}{\theta +1} e^{-\theta x} \Big)^\alpha\biggr]^2},
\ee
for $x > 0$, $\alpha > 0$, $\theta >0$, and $0 < p <1$.

\begin{figure}[!htbp]
\begin{center}
\includegraphics[scale=0.8]{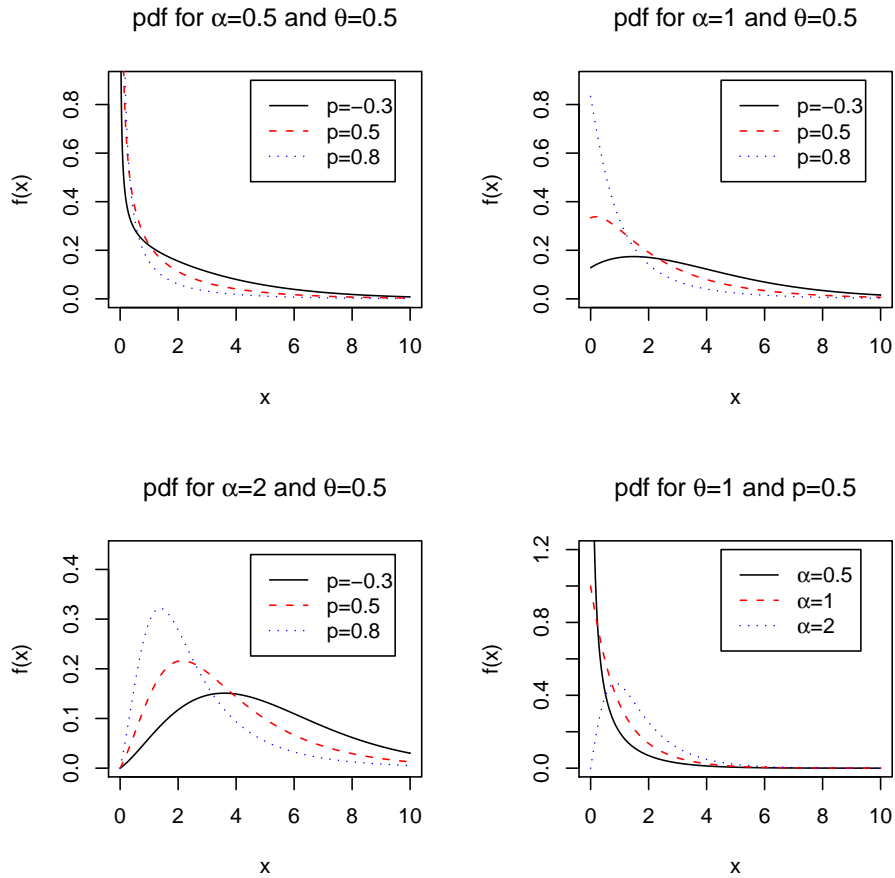}
\end{center}
\caption{Plots of the pdf of the ELG distribution for different values of $\alpha, ~\theta$, and $p$.}
\label{fig:01}
\end{figure}

It should be noticed that even when $p \le 0$, equation (\ref{pdf:01}) is still a well-defined density function, and thus, we can define the ELG distribution in (\ref{pdf:01}) to any $ p < 1$. As mentioned in Section \ref{section:01}, the ELG distribution contains several special submodels. When $\alpha =1$, we have the LG distribution due to \cite{Zake:Mahm:2012}. When $p=0$ and $\alpha =1$, we obtain the Lindley distribution due to \cite{Lind:1958}. The ELG distribution tends to a distribution degenerating at the point 0 when $p \go 1^{-}$.

Figure \ref{fig:01} displays the pdf of the ELG distribution in (\ref{pdf:01}) with selected values of $\alpha, ~\theta$, and $p$. As can be seen in Figure \ref{fig:01}, the shape of the pdf is monotonically decreasing with the modal value of $\infty$ at $x = 0$ when $\alpha <1$ and the shape of the pdf appears upside down bathtub for $\alpha >1$. In particular, when $\alpha=1$, we observe that the shape exhibits monotonically decreasing as well as unimodal, and this observation coincides with Theorem 1 of \cite{Zake:Mahm:2012}, which states that \textit{the density function of the LG distribution is (i) decreasing for all values of $p$ and $\theta$ for which $p > \frac{1-\theta^2}{1+\theta^2}$, (ii) unimodal for all values of $p$ and $\theta$ for which $p \leq \frac{1-\theta^2}{1+\theta^2}$}.

Note also that $f(x) \sim [\alpha\theta(1-p)] e^{-\theta x}$ as $x \go \infty$ and that $f(x) \sim \{\alpha\theta^{\alpha+1}/[(\theta+1)(1-p)]\} x^{\alpha-1}$ as $x \go 0$. Hence, the upper tails of the ELG distribution are exponential, whereas its lower tails are polynomial.

\begin{figure}[!htbp]
\begin{center}
\includegraphics[scale=0.8]{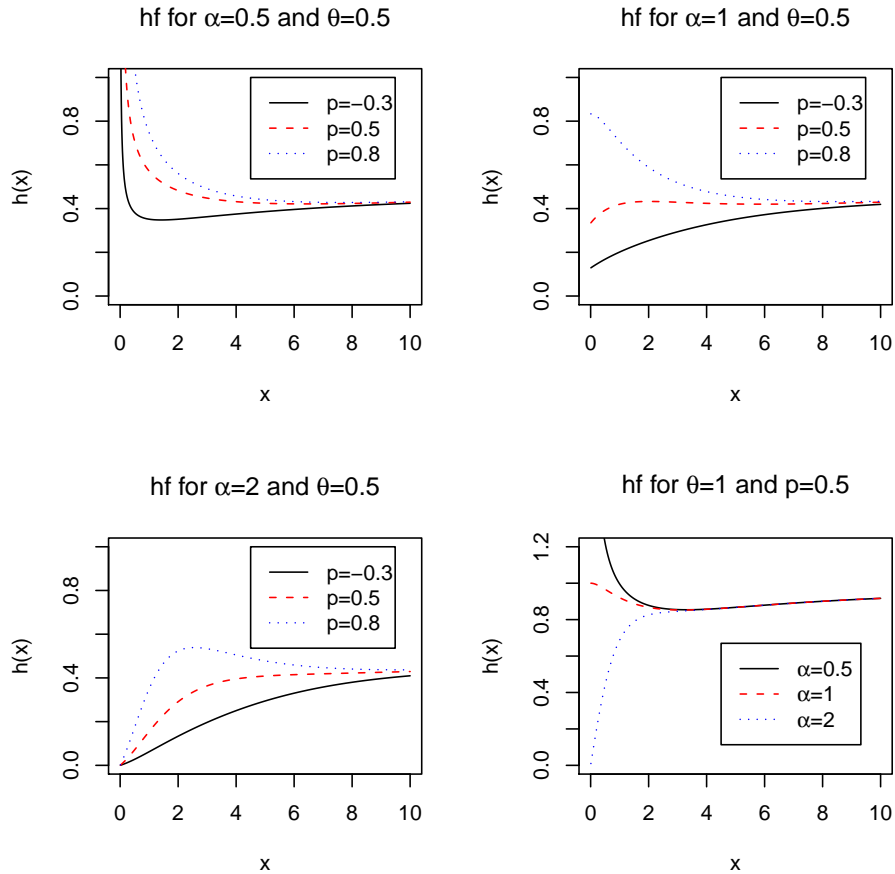}
\end{center}
\caption{Plots of the hazard function (hf) of the ELG distribution for different values of $\alpha, ~\theta$, and $p$.}
\label{fig:02}
\end{figure}

\subsection{Hazard rate function} \label{sec:02}

It is well known that the failure rate function, also known as the hazard rate function, is an important characteristic for lifetime modeling. For a continuous distribution with the cdf $F(x)$ and pdf $f(x)$, the failure rate function is defined as
\begin{equation*}
h(x) = \lim_{ \Delta x \go 0} = \frac{P(X < x+  \Delta x \mid X > x)}{ \Delta x} = \frac{f(x)}{S(x)},
\end{equation*}
where $S(x) = 1 - F(x)$ is the survival function of $X$. Simple algebra provides that the hazard rate function of the ELG distribution is given by
\be \label{hf:01}
h(x)= \frac{\alpha\theta^2(1+x)e^{-\theta x}\Bigl(1 - \frac{\theta +1 + \theta x}{\theta +1} e^{-\theta x} \Bigr)^{\alpha-1}}{(\theta+1)\biggl[1- \Bigl(1 - \frac{\theta +1 + \theta x}{\theta +1} e^{-\theta x} \Bigr)^{\alpha}\biggr]\biggl[1 - p + p\Big(1 - \frac{\theta +1 + \theta x}{\theta +1} e^{-\theta x} \Big)^\alpha\biggr]}
\ee
for $x > 0$, $\alpha > 0$, $\theta >0$, and $p <1$.

Figure \ref{fig:02} depicts possible shapes of equation (\ref{hf:01}) with selected values of $\alpha, ~\theta$, and $p$. It is observed that the hazard function of the new distribution is quite flexible and that the shape of the hazard rate function depends on the values of the three unknown parameters. For example, the shape appears monotonically decreasing if $\alpha$ is sufficiently small and $p$ is not sufficiently large. The shape appears monotonically increasing for small $p$ and large $\alpha$. In addition, the shape may appear bathtub-shaped or firstly increasing then bathtub-shaped for $\alpha =1$. Considering that most real-life systems exhibit different hazard rate shapes: increasing, decreasing, bathtub shaped and unimodal. It has been shown that the ELG distribution exhibits monotone as well as non-monotone hazard rates but does not exhibit a constant hazard rate, which makes the ELG distribution to be superior to many other lifetime distributions having only monotonically increasing, monotonically decreasing, or constant hazard rates.

Note also that as $x \go 0$, the initial hazard rates behave as $h(x) \sim \{\alpha\theta^{\alpha+1}/[(\theta +1)(1-p)]\}x^{\alpha-1}$, which implies that $h(0) \go \infty$ for $\alpha <1$, $h(0) = \alpha\theta^{\alpha+1}/[ (\theta +1)(1-p)]$ for $\alpha = 1$, and $h(0) = 0$ for $\alpha > 1$. The ultimate hazard rate is $h(x) = \theta$ as $x \go \infty$ for all values of $\alpha$.

\subsection{Quantile function} \label{sec:03}

Let $Z$ denote a Lindley random variable with the cdf given by (\ref{lind:cdf}). It can be seen from \cite{Jodr:2010} that the quantile function of the Lindley distribution is given by
\be \label{quant:lind}
G^{-1}(u) = -1 - \frac{1}{\theta} - \frac{1}{\theta}W_{-1}\biggl(-\frac{\theta +1}{e^{\theta+1}}(1-u)\biggr),
\ee
where $0 < u < 1$ and $W_{-1}(\cdot)$ denotes the negative branch of the Lambert $W$ function (i.e., the solution of the equation $W(z) e^{W(z)} = z$). Note that the $W_{-1}(\cdot)$  can be easily computed using the ${lambert\_Wm1}(\cdot)$ function in the package of \textit{gsl} in R language, which is a non-commercial, open-source software package for statistical computation. It can be obtained at no cost from\textit{ http://www.r-project.org}; see \cite{R:2011}.

Let $X$ denote a ELG random variable with the cdf given by (\ref{new:cdf}). By inverting $G(x) = u$ for $0 < u < 1$, we obtain
\begin{equation*}
\biggl(\frac{u - up}{1 -up}\biggr)^{1/\alpha} = 1 - \frac{\theta+ 1+ \theta x}{\theta +1}e^{-\theta x} = G(x).
\end{equation*}
Therefore, it readily follows from equation (\ref{quant:lind}) that the quantile function of the ELG distribution is given
by
\be \label{quand:ELG}
F^{-1}(u) = -1 - \frac{1}{\theta} - \frac{1}{\theta}W_{-1}\biggl(-\frac{\theta +1}{e^{\theta+1}}\biggl[1-\Bigl(\frac{u - up}{1 -up}\Bigr)^{1/\alpha}\biggl]\biggr).
\ee
Note that $-\frac{1}{e} < -\frac{\theta +1}{e^{\theta+1}}\biggl[1-\Bigl(\frac{u - up}{1 -up}\Bigr)^{1/\alpha}\biggl] < 0$, so the $W_{-1}(\cdot)$ is unique, which implies that $F^{-1}(u)$ is also unique. Thus, one can use equation (\ref{quand:ELG}) for generating random data from the ELG distribution. In particular, the quartiles of the ELG distribution, respectively, are given by
\begin{align*}
Q_1 &= F^{-1}\Bigl(\frac{1}{4}\Bigr) = -1 - \frac{1}{\theta} - \frac{1}{\theta}W_{-1}\biggl(-\frac{\theta +1}{e^{\theta+1}}\biggl[1-\Bigl(\frac{1 - p}{4 -p}\Bigr)^{1/\alpha}\biggl]\biggr),\\[+3pt]
Q_2 &= F^{-1}\Bigl(\frac{1}{2}\Bigr) = -1 - \frac{1}{\theta} - \frac{1}{\theta}W_{-1}\biggl(-\frac{\theta +1}{e^{\theta+1}}\biggl[1-\Bigl(\frac{1 - p}{2 -p}\Bigr)^{1/\alpha}\biggl]\biggr),\\[+3pt]
Q_3 &= F^{-1}\Bigl(\frac{3}{4}\Bigr) = -1 - \frac{1}{\theta} - \frac{1}{\theta}W_{-1}\biggl(-\frac{\theta +1}{e^{\theta+1}}\biggl[1-\Bigl(\frac{3 - 3p}{4 -3p}\Bigr)^{1/\alpha}\biggl]\biggr).
\end{align*}

\subsection{Limiting distributions of order statistics} \label{sec:04}

Let $X_1, \cdots, X_n$ be a random sample from the ELG distribution. Sometimes, it would be of interest to study the asymptotic distribution of the extreme values $X_{(1)} = \min\{X_1, \cdots, X_n\}$ and $X_{(n)} = \max\{X_1, \cdots, X_n\}$. By using L'Hospital's rule, we have
\begin{align*}
\lim_{t \go \infty}\frac{1 - F(t + x/\theta)}{1 - F(t)} &=\lim_{t \go \infty}\frac{f(t + x/\theta)}{f(t)}=\frac{1 - \Bigl[1 - \frac{\theta t}{\theta+1}e^{-(\theta t + x)}\Bigr]^\alpha}{1 - \Bigl[1 - \frac{\theta t}{\theta+1}e^{-\theta t}\Bigr]^\alpha}\\
&= e^{-x}.
\end{align*}
In addition, it can also be seen using L'Hospital's rule that
\begin{align*}
\lim_{t \go 0}\frac{F(tx)}{F(t)} &= \lim_{t \go 0}\frac{xf(x t)}{f(t)} = \lim_{t \go 0} \biggl(\frac{1 - \frac{\theta+1+\theta t x}{\theta+1}e^{-\theta t x}}{1 - \frac{\theta+1+\theta t }{\theta+1}e^{-\theta t}}\biggr)^\alpha\\
&=x^\alpha.
\end{align*}
Hence, it directly follows from Theorem 1.6.2 in \cite{Lead:Lind:Root:1983} that there must be norming constants $a_n >0$, $b_n$, $c_n >0$, and $d_n$, such that
\begin{equation*}
Pr\bigl[ a_n(X_{(1)} - b_n) \leq x\bigr] \go \exp\bigl(-e^{-x}\bigr)
\end{equation*}
and
\begin{equation*}
Pr\bigl[ c_n(X_{(n)} - d_n) \leq x\bigr] \go 1 - \exp\bigl(-x^a\bigr)
\end{equation*}
as $n \go \infty$. By following Corollary 1.6.3 in \cite{Lead:Lind:Root:1983}, we can determine the form of the norming constants. As an illustration, one can see that $a_n = \theta$ and $b_n = F^{-1}(1 - 1/n)$, where $F^{-1}(\cdot)$ denotes the inverse function of $F(\cdot)$.


\subsection{Moment properties} \label{sub:mom}

In order to derive the $n$th moment of the ELG distribution, we firstly consider the Taylor series expansion of the form
\begin{equation*}
(1+x)^{-a} = \sum_{k=0}^\infty{{-a \choose k} x^k},
\end{equation*}
for $|x| <1$, which provides that
\begin{align*}
\bigl[1-p + p G^\alpha(x)\bigr]^{-1} &= \frac{1}{1 - p}\Bigl[1 + \frac{p}{1-p}G^\alpha(x)\Bigr]^{-1}\\[+3pt]
&= \frac{1}{1 - p}\sum_{k=0}^\infty{{-1 \choose k}\Bigl[\frac{p}{1-p}G^\alpha(x)\Bigr]^k},
\end{align*}
for $|\frac{p}{1-p}G^\alpha(x)| <1$. Therefore, we can rewrite equation (\ref{new:cdf}) as
\begin{equation*}
F(x) = \frac{1}{1 - p}\sum_{k=0}^\infty{{-1 \choose k}\frac{p^k}{(1-p)^k}\biggl(1 - \frac{\theta +1 + \theta x}{\theta +1} e^{-\theta x} \biggr )^{\alpha k +\alpha}}.
\end{equation*}
We observe that the ELG distribution is a mixture of exponentiated Lindley distributions introduced by \cite{Nada:Bako:2011}. They show that if $Y$ is an exponentiated Lindley random variable with parameters $\theta$ and $\beta$, the $n$th moment and the moment generating function of $Y$ are, respectively, given by
\begin{equation*}
\expect(Y^n_{\theta, \beta}) = \frac{\beta\theta^2}{1 + \theta} K(\beta, \theta, n, \theta)
\end{equation*}
and
\begin{equation*}
M_{Y_{\theta, \beta}}(t) = \frac{\beta\theta^2}{1 + \theta} K (\beta, \theta, 0, \theta-t),
\end{equation*}
where
\begin{align*}
K(a, b, c, \delta) =& \sum_{i=0}^\infty {a-1 \choose i }\frac{(-1)^i}{(1+b)^i}\sum_{j=0}^i {i \choose j} b^j \sum_{k=0}^{j+1} {j+1 \choose k}\\[+3pt]
& \times \int_0^\infty x^{c+k} \exp\bigl( -bix - \delta x\bigr)\, dx.
\end{align*}

By using equation (7), we obtain the $n$th moment of $X$ can be written as
\begin{align}\nonumber \label{expect:01}
\expect(X^n) &= \frac{1}{1 - p}\sum_{k=0}^\infty \frac{(-p)^k}{(1-p)^k}\expect\bigl(Y^n_{\theta, \alpha k + \alpha}\bigr)\\[+3pt]
&=\frac{1}{1 - p}\sum_{k=0}^\infty \frac{(-p)^k}{(1-p)^k}\frac{(\alpha k + \alpha)\theta^2}{1 + \theta} K(\alpha k + \alpha, \theta, n, \theta),
\end{align}
for $n = 1, 2, \cdots$. Equation (\ref{expect:01}) can thus be adopted to compute the mean, skewness, and kurtosis of $X$. The moment generating function of $X$ is
\begin{align*}
M_X(t) &= \frac{1}{1-p}\sum_{k=0}^\infty \frac{(-p)^k}{(1-p)^k} M_{Y_{\theta, \alpha k + \alpha}}(t)\\[+3pt]
&=\frac{1}{1-p}\sum_{k=0}^\infty \frac{(-p)^k}{(1-p)^k} \frac{(\alpha k + \alpha)\theta^2}{1 + \theta} K (\alpha k + \alpha, \theta, 0, \theta-t).
\end{align*}

\section{Estimation of parameters} \label{section:03}

\subsection{Maximum-likelihood estimation} \label{subsection:03}

It is well known that the maximum likelihood estimation procedure is one of the most popular ways for estimating the parameters of continuous distributions because of its attractive properties, such as consistency, asymptotic normality, etc. Let $X_1, \cdots, X_n$ be a random sample from the ELG distribution with unknown parameter vector $\phi = (\theta, \alpha, p)$. Then the log-likelihood function $l = l(\phi; \bfx)$ is given by
\begin{align}\label{lik:01} \nonumber
l = &~n\log\alpha + 2n\log\theta - n\log(\theta+1)+n\log(1-p) + \sum_{i=1}^n{\log(1+x_i)} - \theta \sum_{i=1}^n{x_i} + (\alpha-1) \\[+3pt]
&~\times \sum_{i=1}^n{\log\biggl(1 - \frac{\theta+1+\theta x_i}{\theta+1}e^{-\theta x_i}\biggr)}-2\sum_{i=1}^n{\log\biggl[ 1 - p + p\Big(1 - \frac{\theta +1 + \theta x_i}{\theta +1} e^{-\theta x_i} \Big)^\alpha\biggr]}.
\end{align}
For notational convenience, let
\begin{equation*}
\tau_i(\theta) = 1 - \frac{\theta+1+\theta x_i}{\theta+1}e^{-\theta x_i},
\end{equation*}
for $i = 1, \cdots, n$. The maximum likelihood estimators of the unknown parameters can be obtained by taking the first partial derivatives of equation (\ref{lik:01}) with respect to $\alpha$, $\theta$, and $p$ and putting them equal to 0. We then have the following likelihood equations
\begin{align} \label{der:01}
\frac{\partial l}{\partial \alpha} &= \frac{n}{\alpha} + \sum_{i=1}^n\log\bigl[\tau_i(\theta)\bigr] -2p\sum_{i=1}^n {\frac{\tau_i^\alpha(\theta) \log\bigl[\tau_i(\theta)\bigr]}{1-p+p\tau_i^\alpha(\theta)}},\\[+3pt] \nonumber
\frac{\partial l}{\partial \theta} &= \frac{2n}{\theta} - \frac{n}{\theta+1} - \sum_{i=1}^n{x_i} + \frac{(\alpha-1)\theta}{(\theta+1)^2}\sum_{i=1}^n{\frac{ x_i(2+\theta+\theta x_i + x_i)e^{-\theta x_i}}{\tau_i(\theta)}} -\frac{2\alpha p\theta}{(\theta+1)^2} \\[+3pt] \label{der:02}
&~~~~ \times \sum_{i=1}^n {\frac{\tau_i^{\alpha-1}(\theta) x_i(2+\theta+\theta x_i + x_i)e^{-\theta x_i}}{1-p+p\tau_i^{\alpha}(\theta)}},\\[+3pt]  \label{der:03}
\frac{\partial l}{\partial p} &= -\frac{n}{1-p} + 2\sum_{i=1}^n {\frac{1 - \tau_i^\alpha(\theta)}{1-p+p\tau_i^\alpha(\theta)}}.
\end{align}
Note that the MLEs, respectively $\hat{\alpha}$, $\hat\theta$ and $\hat p$ of $\alpha$, $\theta$ and $p$ cannot be solved analytically through equations (\ref{der:01}), (\ref{der:02}), and (\ref{der:03}). Numerical iteration techniques, such as the Newton-Raphson algorithm, are thus adopted to solve these equations. To employ the Newton-Raphson algorithm, the second derivatives of the log-likelihood are required for all iterations involved in the algorithm. In the following section, an EM algorithm will be given for estimating the maximum likelihood estimates of the unknown parameters.

For interval estimation of the unknown parameters, we consider suitable pivotal quantities based on the asymptotic properties of the MLEs and approximate the distributions of these quantities by the normal distribution. We firstly observe that
\begin{align*}
\frac{ \partial^2 \log l}{\partial\alpha^2} &=-\frac{n}{\alpha^2} - 2p(1-p)\sum_{i=1}^n{\frac{\tau_i^\alpha(\theta)[\log(\tau_i(\theta))]^2} {[1-p+p\tau_i^\alpha(\theta)]^2}},\\[3pt]
\frac{ \partial^2 \log l}{\partial\theta^2} &=-\frac{2n}{\theta^2} + \frac{n}{(\theta+1)^2} -\frac{(\alpha-1)}{(\theta+1)^4}\sum_{i=1}^n{
\frac{x_i e^{-\theta x_i}\big[e^{-\theta x_i}(x_i + 2x_i\theta+2+2\theta)+(t+1)\kappa_i\big]}{\tau_i^2(\theta)} }\\[+3pt]
& ~~~~+\frac{2\alpha p\theta^2}{(\theta+1)^4}\sum_{i=1}^n{\frac{x_i^2(2+\theta+\theta x_i + x_i)^2e^{-2\theta x_i}\tau_i^{\alpha-2} (\theta)\bigl\{(1-\alpha)[1-p+p\tau_i^\alpha(\theta)]+ \alpha p \tau_i^\alpha(\theta)\bigr\}}{\bigl[1-p+p\tau_i^\alpha(\theta)\bigr]^2}}\\[+3pt]
& ~~~~+\frac{2\alpha p}{(\theta+1)^3}\sum_{i=1}^n{\frac{x_i\kappa_i \tau_i^{\alpha-1}(\theta)e^{-\theta x_i}}{1-p+p\tau_i^\alpha(\theta)}},\\[+3pt]
\end{align*}
\begin{align*}
\frac{ \partial^2 \log l}{\partial p^2} &= -\frac{n}{(1-p)^2} + 2\sum_{i=1}^n {\biggl[\frac{1 - \tau_i^\alpha(\theta)}{1-p+p\tau_i^\alpha(\theta)}\biggr]^2},\\[+3pt]
\frac{ \partial^2 \log l}{\partial\alpha\partial\theta} &= \frac{ \partial^2 \log L}{\partial\theta\partial\alpha} =\frac{\theta}{(\theta+1)^2}\sum_{i=1}^n{\frac{x_i(2+\theta+\theta x_i + x_i)e^{-\theta x_i}}{\tau_i(\theta)}}- \frac{2p\theta}{(\theta+1)^2} \\[+3pt]
& ~~~~\times \sum_{i=1}^n{\frac{\bigl[\alpha(1-p)\log(\tau_i(\theta))+ 1-p+p\tau_i^\alpha(\theta)\bigr]x_i(2+\theta+\theta x_i + x_i)e^{-\theta x_i}\tau_i^{\alpha-1}(\theta)}{[1-p+p\tau_i^\alpha(\theta)]^2}},\\[+3pt]
\frac{ \partial^2 \log l}{\partial\alpha\partial p} &= \frac{ \partial^2 \log L}{\partial p\partial\alpha} = - \frac{\tau_i^\alpha(\theta)\log(\tau_i(\theta))}{[1-p+p\tau_i(\theta)]^2},\\[+3pt]
\frac{ \partial^2 \log l}{\partial\theta\partial p } &= \frac{ \partial^2 \log L}{\partial p\partial\theta} =-2\frac{\tau_i^\alpha(\theta)\log(\tau_i(\theta))}{[1-p+p\tau_i^\alpha(\theta)]^2},
\end{align*}
where $\kappa_i = (\theta^3+\theta)(x_i + x_i^2) + \theta^2(3x_i + 2x_i^2)-x_i-2$ for $i=1, \cdots, n$. Then the observed Fisher information matrix of $\alpha$, $\theta$, and $p$ can be written as
\begin{align*}
\bfI = -\left( \begin{array}{ccc} \frac{ \partial^2 \log l}{\partial\alpha^2} & \frac{ \partial^2 \log l}{\partial\alpha\partial\theta} & \frac{ \partial^2 \log l}{\partial\alpha\partial p} \\[+3pt]
\frac{\partial^2 \log l}{\partial\theta\partial\alpha}  & \frac{\partial^2\log l}{\partial\theta^2} & \frac{ \partial^2 \log l}{\partial\theta\partial p } \\[+3pt]
\frac{\partial^2 \log l}{\partial p \partial\alpha}  & \frac{\partial^2\log l}{\partial p \partial \theta} & \frac{ \partial^2 \log l}{\partial p^2 } \end{array} \right),
\end{align*}
so the variance-covariance matrix of the MLEs $\hat\alpha$, $\hat\theta$ and $\hat p$ may be approximated by inverting the matrix $\bfI$ and is thus given by

\begin{equation*}
\bfV = -\left( \begin{array}{ccc} \frac{ \partial^2 \log l}{\partial\alpha^2} & \frac{ \partial^2 \log l}{\partial\alpha\partial\theta} & \frac{ \partial^2 \log l}{\partial\alpha\partial p} \\[+3pt]
\frac{\partial^2 \log l}{\partial\theta\partial\alpha}  & \frac{\partial^2\log l}{\partial\theta^2} & \frac{ \partial^2 \log l}{\partial\theta\partial p } \\[+3pt]
\frac{\partial^2 \log l}{\partial p \partial\alpha}  & \frac{\partial^2\log l}{\partial p \partial \theta} & \frac{ \partial^2 \log l}{\partial p^2 } \end{array} \right)^{-1} =
\left( \begin{array}{ccc} \var(\alpha)  &\cov(\alpha, \theta)  &\cov(\alpha, p) \\[+3pt]
\cov(\theta, \alpha)   & \var(\theta)   &\cov(\theta, p)\\[+3pt]
\cov(p, \alpha)   & \cov(p, \theta)   &\var(p)
\end{array} \right).
\end{equation*}
The asymptotic joint distribution of the MLEs $\hat\alpha$, $\hat\theta$, and $\hat p$ is then approximately multivariate normal and is given by
\be \label{asym:normal}
\left(
\begin{array}{c}
\hat\alpha\\
\hat\theta\\
\hat p
\end{array}
          \right)
\sim
N\left[ \left(\begin{array}{c}
\alpha\\
\theta\\
p
\end{array}\right)
, ~~\left( \begin{array}{ccc} \var(\alpha)  &\cov(\alpha, \theta)  &\cov(\alpha, p) \\[12pt]
\cov(\theta, \alpha)   & \var(\theta)   &\cov(\theta, p)\\[12pt]
\cov(p, \alpha)   & \cov(p, \theta)   &\var(p)
\end{array} \right) \right].
\ee
Noting that $\bfV$ involves the unknown parameters $\alpha$, $\theta$, and $p$, we replace these parameters by their corresponding MLEs to obtain an estimate of $\bfV$ denoted by
\begin{equation*}
\widehat{\bfV} = \left(\begin{array}{ccc} \widehat{\var(\alpha)}  & \widehat{\cov(\alpha, \theta)}   & \widehat{\cov(\alpha, p)} \\ [12pt]
\widehat{\cov(\theta,\alpha)}   & \widehat{\var(\theta)}  & \widehat{\cov(\theta, p)} \\ [12pt]
\widehat{\cov(p,\alpha)}   & \widehat{\cov(p, \theta)}  & \widehat{\var(p)}
\end{array}\right).
\end{equation*}
Thus, by using (\ref{asym:normal}), the asymptotic $100(1-\gamma)\%$ confidence intervals of $\alpha$, $\theta$, and $p$ are determined by
\begin{align*}
&\Bigl[ \hat\alpha - z_{\gamma/2}\sqrt{\widehat{\var(\alpha)}}, ~\hat\alpha + z_{\gamma/2}\sqrt{\widehat{\var(\alpha)}}\Bigr],\\[+3pt]
&\Bigl[ \hat\theta - z_{\gamma/2}\sqrt{\widehat{\var(\theta)}}, ~\hat\theta + z_{\gamma/2}\sqrt{\widehat{\var(\theta)}}\Bigr],\\[+3pt]
&\Bigl[ \hat p - z_{\gamma/2}\sqrt{\widehat{\var(p)}}, ~\hat p + z_{\gamma/2}\sqrt{\widehat{\var(p)}}\Bigr],
\end{align*}
respectively, where $z_p$ is the upper $p$th percentile of the standard normal distribution.

It deserves to mention that the likelihood ratio (LR) can be used to evaluate the difference between the ELG distribution and its special submodels. We partition the parameters of the ELG distribution into $(\phi_1', \phi_2')'$, where $\phi_1$ is the parameter of interest and $\phi_2$ is the remaining parameters. Consider the hypotheses
\begin{equation} \label{test:01}
H_0: \phi_1 = \phi_1^{(0)} \quad \mathrm{versus} \quad H_1: \phi_1 \neq \phi_1^{(0)}.
\end{equation}
Suppose we are interested in testing the null hypothesis $H_0$ against the alternative hypothesis $H_1$. The LR statistic for the test of the null hypothesis in (\ref{test:01}) is
\begin{equation} \label{test:02}
\omega = 2\bigl\{ l(\hat\phi; \bfx) - l(\hat\phi^\ast; \bfx)\bigr\},
\end{equation}
where $\hat\phi$ and $\hat\phi^\ast$ are the restricted and unrestricted maximum likelihood estimators under the null and alternative hypotheses, respectively. Under the null hypothesis,
\be \label{test:03}
\omega \stackrel{D}{\longrightarrow} \chi^2_\kappa,
\ee
where $\stackrel{D}{\longrightarrow}$ denotes convergence in distribution as $n \go \infty$ and $\kappa$ is the dimension of the subset $\phi_1$ of interest. For instance, we can compare the ELG and LG distributions by testing $H_0: \alpha=1$ versus $H_1: \alpha \neq 1$. The ELG and Lindley distributions are compared by testing $H_0: (\alpha, p) =(1,0)$ versus $H_1: (\alpha, p)  \neq (1,0)$.

\subsection{Expected-maximization algorithm}

\cite{Demp:Lair:Rubi:1977} introduce a general iterative approach, the so-called EM algorithm, as a very powerful tool for estimating the parameters in cases where observations are treated as incomplete data. Suppose that $X = (X_{1}, X_{2}, \cdots, X_{n})$ and $Z = (Z_1, Z_2, \cdots, Z_{n})$ represent the observed and hypothetical data, respectively. Here, the hypothetical data can be thought of as missing data because $Z_1, Z_2, \cdots, Z_{n}$ are not observable. In this paper, we can formulate the problem of finding the MLEs of the unknown parameters as an incomplete data problem, and thus, the EM algorithm is applicable to determine the MLEs of the ELG distribution. Let $W = (X, Z)$ denote the complete data. To start this algorithm, define the pdf of for each $(X_i, Z_i)$ for $i = 1, \cdots, n$ in the form
\begin{align*}
g(x, z, \alpha, \theta, p) &= \frac{\alpha(1-p)\theta^2 z(1+x)}{\theta+1}e^{-\theta x}\biggl(1 - \frac{\theta+1+\theta x}{\theta+1}e^{-\theta x} \biggr)^{\alpha-1}\\[+3pt]
&~~~~ \times \biggl[p - p\Bigl(1-\frac{\theta+1+\theta x}{\theta+1}e^{-\theta x}\Bigr)^\alpha\biggr]^{z-1}.
\end{align*}
The E-step of an EM cycle requires the conditional expectation of $(Z\mid X, \alpha^{(r)}, \theta^{(r)}, p^{(r)})$, where $(\alpha^{(r)}, \theta^{(r)}, p^{(r)})$ is the current estimate of $(\alpha, \theta, p)$ in the $r$the iteration. Note that the pdf of $Z$ given $X$, say $g(z \mid x)$, is given by
\begin{equation*}
g(z \mid x) = \frac{z\biggl[p - p\Bigl(1-\frac{\theta+1+\theta x}{\theta+1}e^{-\theta x}\Bigr)^\alpha\biggr]^{z-1}}{\biggl[1-p+p\Bigl(1-\frac{\theta+1+\theta x}{\theta+1}e^{-\theta x}\Bigr)^\alpha\biggr]^2}.
\end{equation*}
Thus, the conditional expectation is given by
\begin{equation*}
\expect[Z \mid X, \alpha, \theta, p] = \frac{1 + p\biggl[1 - \Bigl(1-\frac{\theta+1+\theta x}{\theta+1}e^{-\theta x}\Bigr)^\alpha\biggr]}{1 - p\biggl[1 - \Bigl(1-\frac{\theta+1+\theta x}{\theta+1}e^{-\theta x}\Bigr)^\alpha\biggr]}.
\end{equation*}
The log-likelihood function $l_c(W; \alpha, \theta, p)$ of the complete data after ignoring the constants can be written as
\begin{align} \label{pse:01}\nonumber
l_c(W; \alpha, \theta, p) &\propto \sum_{i=1}^n{z_i} + n \log\alpha + \sum_{i=1}^n{\log(1+x_i)} + 2n\log\theta -n \log(\theta+1) \\[+3pt] \nonumber
&~~~ - \theta\sum_{i=1}^n{x_i} + n\log(1-p)+ (\alpha-1)\sum_{i=1}^n{\log\biggl(1-\frac{\theta+1+\theta x_i}{\theta+1}e^{-\theta x_i}\biggr)} \\[+3pt]
&~~~+ \sum_{i=1}^n{(z_i-1)\log\biggl[p - p\Bigl(1-\frac{\theta+1+\theta x_i}{\theta+1}e^{-\theta x_i}\Bigr)^\alpha\biggr]}.
\end{align}
Next the M-step involves the maximization of the pseudo log-likelihood function (\ref{pse:01}). The components of the score function are given by
\begin{align*}
\frac{\partial l_c}{\partial \alpha} &= \frac{n}{\alpha} + \sum_{i=1}^n{\log\Bigl(1-\frac{\theta+1+\theta x_i}{\theta+1}e^{-\theta x_i}\Bigr)}- \sum_{i=1}^n{(z_i-1)\frac{\Bigl(1-\frac{\theta+1+\theta x_i}{\theta+1}e^{-\theta x_i}\Bigr)^\alpha\log\Bigl(1-\frac{\theta+1+\theta x_i}{\theta+1}e^{-\theta x_i}\Bigr)}{1 - \Bigl(1-\frac{\theta+1+\theta x_i}{\theta+1}e^{-\theta x_i}\Bigr)^\alpha}},\\[+3pt] \nonumber
\frac{\partial l_c}{\partial \theta} &=\frac{2n}{\theta} - \frac{n}{\theta+1} - \sum_{i=1}^n{x_i} + (\alpha-1)\sum_{i=1}^n{\frac{\theta x_i e^{-\theta x}\bigl(1 + x_i + \frac{1}{\theta+1}\bigr)}{(\theta+1)\Bigl(1-\frac{\theta+1+\theta x_i}{\theta+1}e^{-\theta x_i}\Bigr)}}-\frac{\alpha\theta}{(\theta+1)^2}\\[+3pt]
&~~~\times\sum_{i=1}^n{\frac{(z_i-1)x_i(2 + \theta + \theta x_i + x_i)e^{-\theta x_i}\Bigl(1-\frac{\theta+1+\theta x_i}{\theta+1}e^{-\theta x_i}\Bigr)^{\alpha-1}}{1- \Bigl(1-\frac{\theta+1+\theta x_i}{\theta+1}e^{-\theta x_i}\Bigr)^\alpha}},\\[+3pt]
\frac{\partial l_c}{\partial p} &= -\frac{n}{1-p} + \sum_{i=1}^n{\frac{z_i-1}{p}}.
\end{align*}
For notational convenience, let
\begin{equation*}
\tau_i^{(r)}(\theta) = 1 - \frac{\theta^{(r)}+1+\theta^{(r)} x_i}{\theta^{(r)}+1}e^{-\theta^{(r)} x_i},
\end{equation*}
for $i = 1, \cdots, n$. By replacing the missing $Z$'s with their conditional expectations $\expect[Z \mid X, \alpha^{(r)}, \theta^{(r)}, p^{(r)}]$, we obtain an iterative procedure of the EM algorithm given by the following equations.
\begin{align*}
&0  = \frac{n}{\alpha^{(r+1)}} + \sum_{i=1}^n{\log\bigl(\tau_i^{(r+1)}(\theta)\bigr)}- \sum_{i=1}^n{(z_i-1)\frac{\bigl(\tau_i^{(r+1)}(\theta)\bigr)^{\alpha^{(r+1)}}\log\bigl(\tau_i^{(r+1)}(\theta)\bigr)}{1 - \bigl(\tau_i^{(r+1)}(\theta)\bigr)^{\alpha^{(r+1)}}}},\\[+3pt] \nonumber
&0  = \frac{2n}{\theta^{(r+1)}} - \frac{n}{\theta^{(r+1)}+1} - \sum_{i=1}^n{x_i} + (\alpha^{(r+1)}-1)\sum_{i=1}^n{\frac{\theta^{(r+1)} x_i e^{-\theta^{(r+1)} x_i}\bigl(1 + x_i + \frac{1}{\theta^{(r+1)}+1}\bigr)}{(\theta^{(r+1)}+1)\tau_i^{(r+1)}(\theta)}}\\[+3pt]
 &~~~{}-\frac{\alpha^{(r+1)}\theta^{(r+1)}}{(\theta^{(r+1)}+1)^2}\sum_{i=1}^n{\frac{(z_i-1)x_i(2 + \theta^{(r+1)} + \theta^{(r+1)} x_i + x_i)e^{-\theta^{(r+1)x_i}}\bigl(\tau_i^{(r+1)}(\theta)\bigr)^{\alpha^{(r+1)}-1}}{1- \bigl(\tau_i^{(r+1)}(\theta)\bigr)^{\alpha^{(r+1)}}}},\\[+3pt]
&p^{(r+1)}  = 1 - \frac{n}{\sum_{i=1}^n{z_i}},
\end{align*}
where
\begin{equation*}
z_i = \frac{1 + p^{(r)}\biggl[1 - \bigl(\tau_i^{(r)}(\theta)\bigr)^{\alpha^{(r)}}\biggr]}{1 - p^{(r)}\biggl[1 - \bigl(\tau_i^{(r)}(\theta)\bigr)^{\alpha^{(r)}}\biggr]},
\end{equation*}
for $i = 1, \cdots, n$. Note that some efficient numerical methods, such as the Newton-Raphson algorithm, are needed for solving the first two equations above.

\section{Application} \label{section:04}

In this section we illustrate the applicability of the ELG distribution by considering a real dataset. The dataset is taken from \citeauthor{Gros:Clar:1975} (\citeyear{Gros:Clar:1975}, p. 105) and shows the relief times of 20 patients receiving an analgesic. The data are presented in Table \ref{table:EX01}. We fit the ELG, Gamma, Weibull, and LG distributions to the real dataset. Namely,
\begin{enumerate}
\item[(i)] Gamma$(\beta, \alpha)$
\begin{align*}
f_1(x) &= \frac{1}{\Gamma(\beta)}\alpha^\beta x^{\beta-1}e^{-\alpha x}, \ \  x > 0, \ \ \beta, \alpha > 0;
\end{align*}
\item[(ii)] Weibull$(\beta, \lambda)$
\begin{align*}
f_2(x) &= \frac{\alpha}{\beta}\biggl(\frac{x}{\beta}\biggr)^{\alpha-1}e^{-(x/\beta)^k},\ \  x > 0, \ \  \beta, \alpha > 0;
\end{align*}
\item[(iii)] LG$(\theta, p)$
\begin{align*}
f_3(x) &= \frac{\theta^2}{\theta+1}(1-p)(1+x)e^{-\theta x}\biggl[1 - \frac{p(\theta+1+\theta x) }{\theta+1}e^{-\theta x}\biggr]^{-2}, \ \  x > 0, \ \  \theta > 0, p < 1.
\end{align*}
\end{enumerate}

\begin{figure}[!htpb]
\begin{center}
\includegraphics[scale=0.5]{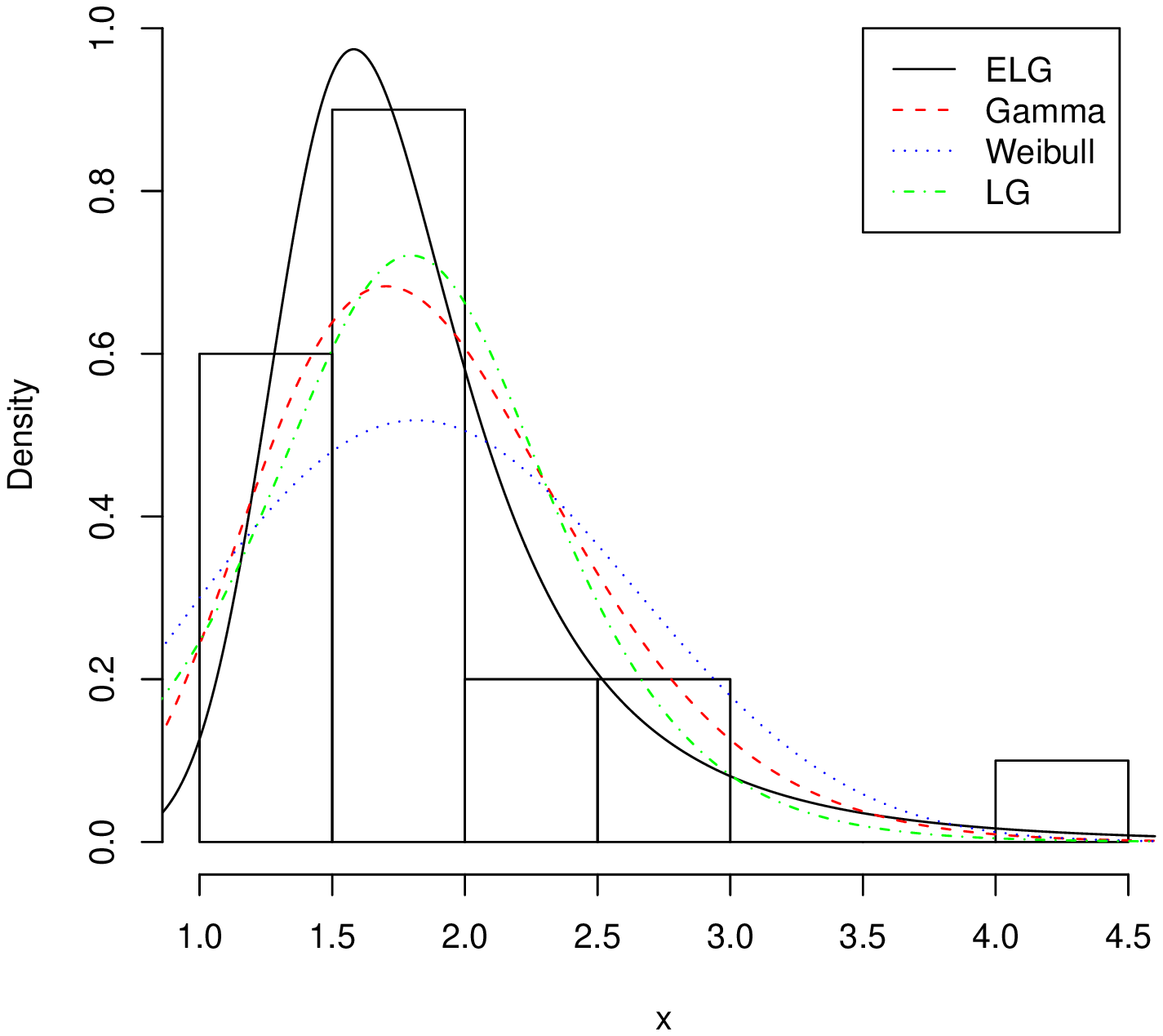}
\includegraphics[scale=0.5]{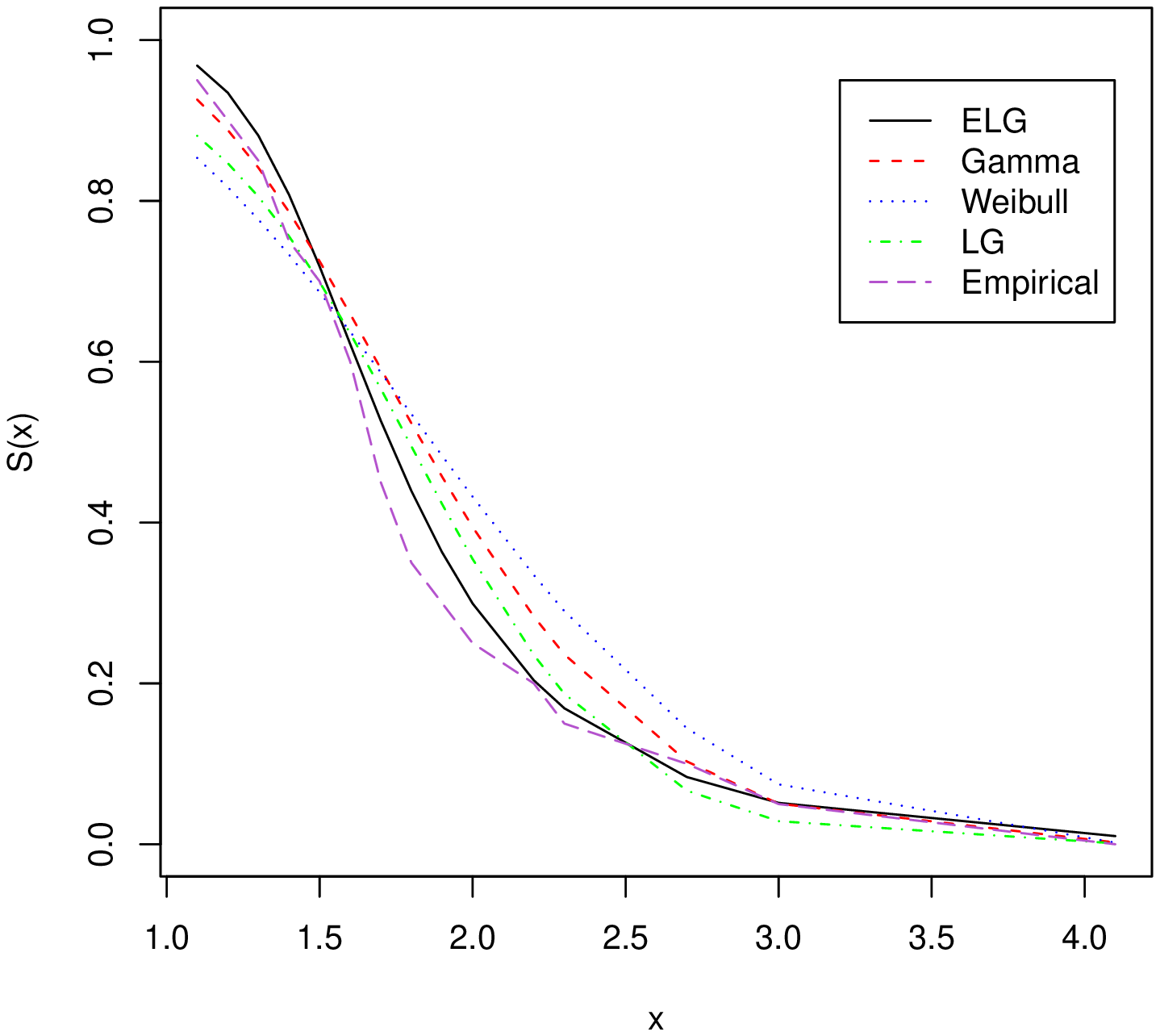}
\end{center}
\caption{Plots of the estimated density and survival functions for the fitted models.}
\label{fig:error002}
\end{figure}

The Akaike information criterion (AIC), the Bayesian information criterion (BIC), and the AIC with a correction (AICc) are advocated to compare the candidate distributions. Table \ref{table:EX02} shows the MLEs of the parameters, AIC, BIC, and AICc for the ELG, Gamma, Weibull, and LG distributions. As can be seen in Table \ref{table:EX02}, the smallest value of each criterion mentioned above is obtained for the ELG distribution only, indicating that the ELG distribution is a strong competitor to other distributions that are commonly used for fitting lifetime data. The plots of the fitted probability density and survival functions are also shown in Figure \ref{fig:error002}. It can be seen from the two figures that the ELG distribution appears to capture the general pattern of the histograms best and that the ELG survival function fits the empirical survival better than the Gamma, Weibull, and LG survival functions.

\vskip4mm
\begin{table}[!htbp]
\centering 
\begin{tabular}{cccccccccccccccccccc}   \hline  \hline
1.1 && 1.4 &&  1.3 &&  1.7 &&  1.9 && 1.8 && 1.6 && 2.2 && 1.7 && 2.7  \\
4.1 && 1.8 &&  1.5 &&  1.2 &&  1.4 && 3.0 && 1.7 && 2.3 && 1.6 && 2.0  \\ \hline \hline
\end{tabular}
\caption{Relief times of twenty patients given by \citeauthor{Gros:Clar:1975} (\citeyear{Gros:Clar:1975}, p. 105).}
\label{table:EX01}
\end{table}

As mentioned in Section \ref{subsection:03}, we can adopt the LR statistic to compare between the ELG distribution and its special submodels. For example, note that the LR statistic for testing between the LG and ELG distributions (i.e., $H_0: \alpha= 1$ versus $H_1 : \alpha \neq 1$) is $\omega = 7.5667$ and the corresponding p-value is $0.0059$. Thus, at the 1\% significance level, there exists significant evidence to reject $H_0$ in favor of the ELG distribution. The same conclusion occurs when testing between the Lindley and ELG distributions. In summary, the ELG distribution improves significantly on the fits of the LG and Lindley distributions.

\vskip4mm
\begin{table}[!htbp]
\centering 
\begin{tabular}{l|lll|lll}   \hline  \hline
Model        &                       &  Parameters          &                       &  AIC    & BIC     & AICc \\ \hline
Gamma        & $\hat\alpha$ = 5.0887 & $\hat\beta = 9.6685$ &                       & 39.6372 & 41.6287 & 40.3431   \\
Weibull      & $\hat\alpha$ = 2.7870& $\hat\beta$ = 2.1300  &                       & 45.1728 & 47.1643 & 45.8787  \\
LG           & $\hat\theta$ = 3.1827 & $\hat p = -125.1293$ &                       & 42.6723 & 44.6638 & 43.3782  \\
ELG          & $\hat\alpha$ = 15.5628& $\hat\theta$ = 1.5270&  $\hat p$ = 0.9059    & 37.1056 & 40.0928 & 38.6056 \\
\hline \hline
\end{tabular}
\caption{MLEs of the fitted models, AIC, BIC, and AICc for the relief times data by \cite{Gros:Clar:1975}.}
\label{table:EX02}
\end{table}

\section{Concluding remarks} \label{section:05}

In this paper, we introduce a new three-parameter distribution, so-called the exponentiated Lindley geometric distribution, that generalizes the LG distribution due to \cite{Zake:Mahm:2012} and the Lindley distribution proposed by \cite{Lind:1958}. many of its standard properties are discussed in detail. These include shape of the probability density function, hazard rate function and its shape, quantile function, limiting distributions of order statistics, and the $n$th moments. Moreover, the maximum likelihood estimation procedure is discussed and an EM algorithm is provided for estimating the parameters. As can be seen from the shapes of the probability density and hazard functions, the new distribution provides more flexibility than other distributions that are commonly used for fitting lifetime data. Finally, a real-data example is analyzed to show the applicability of the new distribution in practical situations.

It deserves to mention that other attractive properties of the new distribution are not considered in this paper, such as the stochastic orderings, cumulants, cumulative residual entropy, R\'{e}nyi and Shannon entropies, distribution of the ratio of ELG random variables, and multivariate generalizations of the ELG distribution, etc. In an ongoing project, we plan to address some of these properties listed above. In this paper, the unknown parameters of the new distribution have only been estimated by the maximum likelihood estimation procedure. Bayesian estimates of the parameters are currently under investigation and will be reported elsewhere.

\bibliographystyle{annals}

\end{document}